\documentclass[preprint,authoryear,3p,times,12pt]{elsarticle}
\usepackage[utf8]{inputenc}
\usepackage{amsfonts,amsmath,bm,bbm}
\usepackage{amsmath}
\usepackage{graphicx}
\usepackage{epstopdf}
\usepackage{amssymb}
\usepackage{mathtools}
\usepackage{enumerate}
\usepackage{natbib}
\usepackage{indentfirst}
\usepackage[font=footnotesize]{caption}
\usepackage{floatrow}
\usepackage{enumitem}
\usepackage{dsfont}

\usepackage{subcaption}

\usepackage{booktabs}
\usepackage{hyperref}

\usepackage{color}
\definecolor{gray}{rgb}{0.5,0.5,0.5}
\definecolor{dgreen}{rgb}{0,0.5,0}

\begin{document}
	\begin{frontmatter}
		
		\title{A portfolio management of a small RES utility with a Structural Vector Autoregressive  model of German electricity markets}

		\author[KBO]{Katarzyna Maciejowska}
		\ead{katarzyna.maciejowska@pwr.edu.pl}
		
		\address[KBO]{Department of Operations Research and Business Intelligence, Wroc{\l}aw University of Science and Technology, 50-370 Wroc{\l}aw, Poland}
		
		\date{This version: \today}

		\begin{abstract}
		 The changes in electricity markets expose RES producers and electricity traders to various risks, among which the price and the volume risk play a very important role. In this research, a portfolio building strategies are presented, which allow to dynamically choose a proportion of electricity traded in different electricity markets (day-ahead and intraday) and hence to optimize the behavior of an utility. Two types of approaches are considered: simple, assuming that the proportions are fixed, and data driven, which allows for thier fluctuation. In order to explore the market information, Structural Vector Autoregressive (SVAR) model is applied, which allows to estimate the relationship between variables of interest and to simulate their future distribution. The presented methods are evaluated with data coming from German electricity market. The results indicate that data driven trading strategies allow to increase the utility revenue and at the same time reduce the trading risk, measured by the predictability of the next day income and the revenue Value at Risk. It turns out that the approach based on Sharp Ratio provides the most robust results.
			
		\end{abstract}

		\begin{keyword}
Intraday electricity market, Day ahead electricity market,  Structural Vector Autoregressive model, probabilistic forecasting,
Trading strategy
		\end{keyword}
		
	\end{frontmatter}

\section{Introduction}

In recent decades, thanks to development of short-term electricity markets, new trade opportunities have opened for generation utilities and demand units. Their operation is no longer optimized centrally and power plants managers act in the market to maximize utility profits. Market participants can now choose, whether to trade via the organized power exchanges, such as Nord Pool or EEX in Europe and PJM in the USA, broker platform or to sign over-the-counter (OTC) contracts.   Although trade in  power exchanges is voluntary, market parties are encouraged to self-balance their positions. This means that no electricity should be left intentionally for the trade on the balancing market (\cite{Koch:Maskos:2020}, \cite{Pape:Hagemann:Weber:2016}).  
Self-balancing is particularly valid for RES utilities, which generation is based on intermittent energy sources, such as wind and solar (\cite{Gianfreda:Parisio:Pelagatti:2016},\cite{Kiesel:Paraschiv:2017}, \cite{Koch:Hirth:2019}).

The changes of electricity markets have been motivated and accompanied by 
dynamic development of RES generation. In the year 2020, RES accounts for 50.5\% of generation in Germany. The increase of RES penetration would not be possible without a variety of support schemes introduced by European countries. First, RES generation is granted a priority during the dispatch, ensuring that all green electricity is efficiently traded. Second, different financial incentives, starting with Feed-in-Tariffs (FIT) have been proposed to increase the profitability of RES investments. In FIT, RES generators are paid a fixed price at a guaranteed level (irrespective of the wholesale price) for the electricity produced and fed into the grid. In Germany, the 2000 Renewable Energy Act guaranteed FIT for wind and solar generation for 20 years. For many installations, this two-decade guarantee is  going to expire soon. As the result, RES producers will need to sell their generation at market prices either via the power exchange or bilateral contracts. 
At the same time, one could observe a general change in the approach toward support schemes and shift from FIT to Feed-in-Premium (FIP) mechanism. In case of FIP, producers receive a premium price, which is a payment (EURO/MWh) in addition to the wholesale price. This premium can be either fixed (Denmark, Lithuania) or floating (Germany, Greece and Netherlands among others). It should be noticed that the ongoing changes leads to a closer relation between revenues of RES generators and wholesale electricity prices. They encourage financially sustainable investments, which respond to market incentives. 

As the result, RES producers become more exposed to various market risks. 
As stated by \cite{Spodniak:etal:2021}, the major ones are price and volume risks. The price risk reflects the fact that electricity prices are stochastic and depends on the unknown future levels of demand and generation structure (\cite{Weron2014}, \cite{Uniejewski:etal:2019}, \cite{Hong:etal:2020}). Additionally, electricity producers face now a cascade of trade opportunities, which includes different markets (bilateral, day-ahead, intrday) and contract types.
From the perspective or RES utility, the volume risk can be analyzed at two levels: a state wise and an individual. The uncertainty about the (individual) utility production stems from intermittent weather conditions, which change continuously up to the delivery time. At the same time, the aggregated generation and consumption volumes are affected by a wider range of  factors, which  include both weather conditions, social events, trading strategies and conventional power plant outages. These two types of risks, price and volume, are closely connected with each other. In particular, the aggregated volume risk impacts the variability of electricity prices. This property has recently attracted attention and has been discussed in the literature (see \cite{Ketterer2014}, \cite{Rintamaki:etal:2017}, \cite{Maciejowska:2020} among others). The research indicates that wind and solar forecast errors impacts both the variance and the whole distribution of electricity prices and are one of the major factors influencing the spread between the day-ahead and intraday prices (\cite{Kiesel:Paraschiv:2017}, \cite{Spodniak:etal:2021}). 

The exposure to price and volume risks leads to a rise of income uncertainty and hence increase the need for appropriate risk management.  The reduction of the revenue risk can be obtained in various ways.
\cite{Kath:etal:2020} show that generators can sign a contract with a trading company, which will allow them to sell all the produced electricity at, for example, day-ahead price and therefore limit its trade risk. On the other hand, \cite{Maciejowska:Nitka:Weron:2019} demonstrate that generators can reduce their price uncertainty by an active trade on two markets: day-ahead and intraday. The results indicate that model based choice of the market, which offers a higher price, can increase the revenues of the utility and reduce its risk. Similarly, \cite{Kath:Ziel:2018} show that the choice between different types of contracts in an intraday market (continuous vs. auctions) can be profitable and lead to considerable financial gains.

Although the portfolio management  seems to be of great importance for practitioners, it has not been studied much in the electricity market literature. Most of the articles address only one source of risk, price or volume. An exception is a paper of \cite{Faria:Fleten:2011}, which proposes a model of bidding strategy for a hydro power plant, which takes into account the stochastic nature of both market prices and generation.  Therefore, the main goal of this research is to fill the existing literature gap. In this paper, Structural Vector Autoregressive (SVAR) model is proposed, which allows to analyze jointly different types of risk and hence considers the input uncertainty \citep[][]{cor:etal:20}.  Four sources of uncertainty are considered: weather conditions, demand shocks and unpredictable behavior of market participants in day-ahead and intraday markets (called speculative shocks). It is shown that the SVAR model can be used for forecasting and simulating of the next day revenue distribution. The outcomes are finally  employed for selection of the optimal portfolio weights. Similar to \cite{Maciejowska:Nitka:Weron:2019}, the resulting trade strategies are evaluated with two types of financial measures: average revenue and associated Value at Risk (VaR).

Vector Autoregressive (VAR) models, although widely used in time series analysis, has not been explored much in modeling electricity markets. \cite{Silva:etal:2012} apply Structural VAR (SVAR) to analyze the relationship between the economic growth and electricity prices.   \cite{Bernstein:Madlener:2015} used yearly data to build a Vector Error Correction model (VECM) to asses the price elasticity of electricity. In both articles, the macroeconomic approach is adopted and low frequency data are analyzed. The higher frequency information, with hourly or daily resolution, has been explored by \cite{Maciejowska2014}, \cite{Paschen:2016}, \cite{Spodniak:etal:2021}. \cite{Spodniak:etal:2021} use VAR models to asses the relevance of different short-term markets, such as day-ahead, intraday and balancing market. They show that due to an increase of the wind power share in the generation mix, the markets closer to the delivery are becoming more important. In \cite{Maciejowska2014} and \cite{Paschen:2016} the impact of different market shocks on day-ahead electricity prices is described with SVAR model. \cite{Paschen:2016} uses the estimates of SVAR model to obtain impulse response functions and to analyze dynamic interrelations between spot prices and RES power. Finally, \cite{Maciejowska2014} shows that speculative shocks, defined as an unpredictable behavior of electricity traders, have the largest share in the electricity price variance.  Up to my knowledge, in the literature there are not publications exploring the potential usage of SVAR model in the decision process of an electricity generator.

The article is structured as follows. Section \ref{sec:Data} describes briefly the data used in the analysis. In Section \ref{Sec:SVAR}, a SVAR model of electricity market is presented, which is next applied to predict a revenue distribution and to support the decision process of a RES utility, Section \ref{Sec:Decision}. Section \ref{sec:Results} presents the results of the experiment and a statistical comparison of performance of proposed trading strategies. Finally, Section \ref{Sec:Conclusions} concludes. 

\section{Data}\label{sec:Data}

In this research, German electricity market is considered. The analysis are based on data published by TSOs and EPEX exchange and cover the period from 01-Oct-2015 to 31-Sep-2019. Since Austria separated from the German the bidding zone, only the data on German generation level and structure is used. The data sets consist of day-ahead ($DA_{th}$) and intraday ($ID_{th}$) prices, with the later one being described by an ID3 index (volume weighted prices from the last 3 hours of trade). The electricity prices are complemented by information on actual levels and system forecasts of fundamental variables: the total load ($L_{th}$) and RES generation ($RES_{th}$) generation. In the remaining part of the paper, the index $h$ stands for an hour and $t$ for a day.

\begin{figure*}[h]
	\centering	

	\begin{subfigure}{0.45\textwidth} 
		\includegraphics[width=\textwidth]{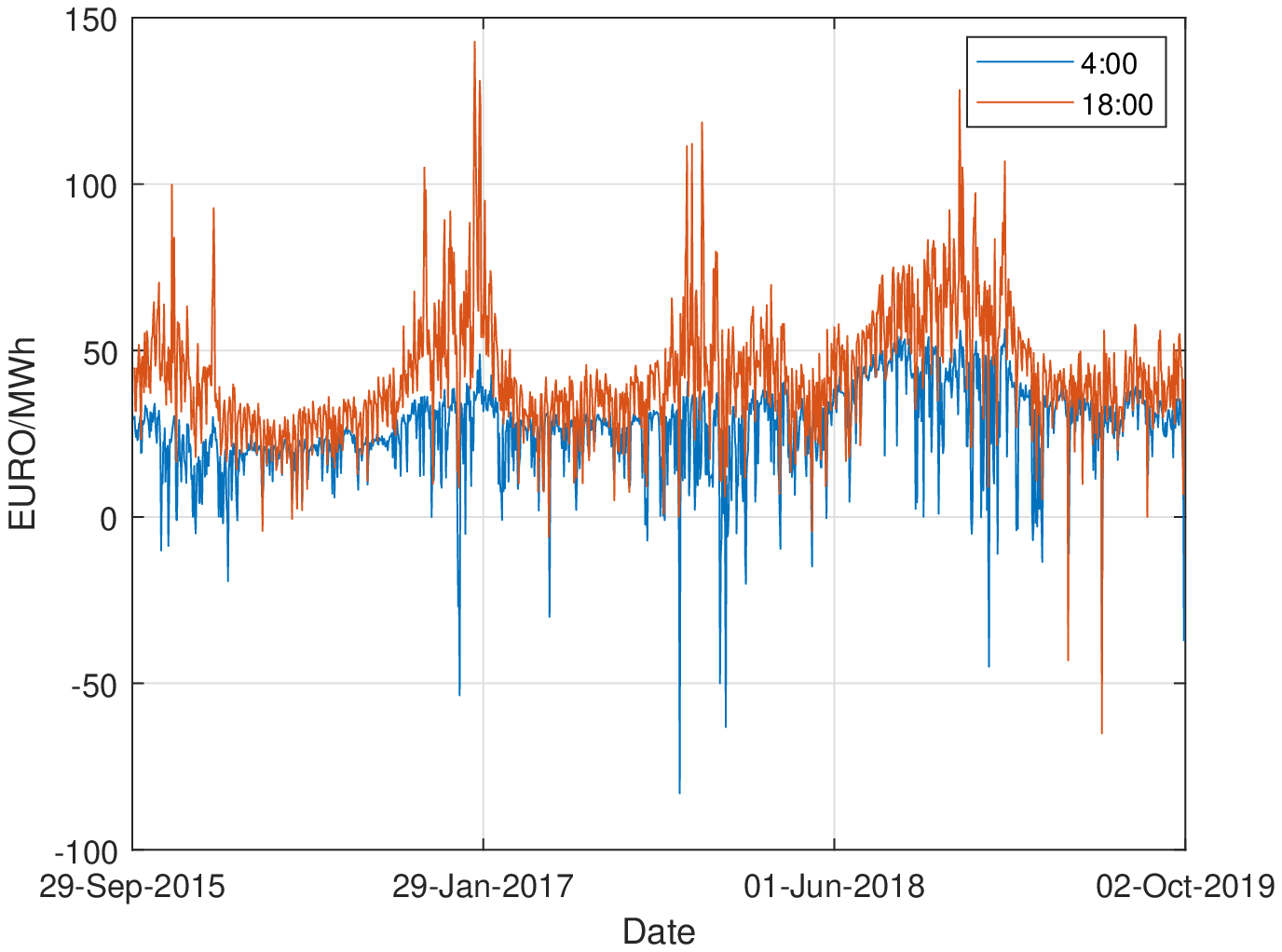}
		\caption{Day-ahead prices}
	\end{subfigure}
	~
	\begin{subfigure}{0.45\textwidth} 
		\includegraphics[width=\textwidth]{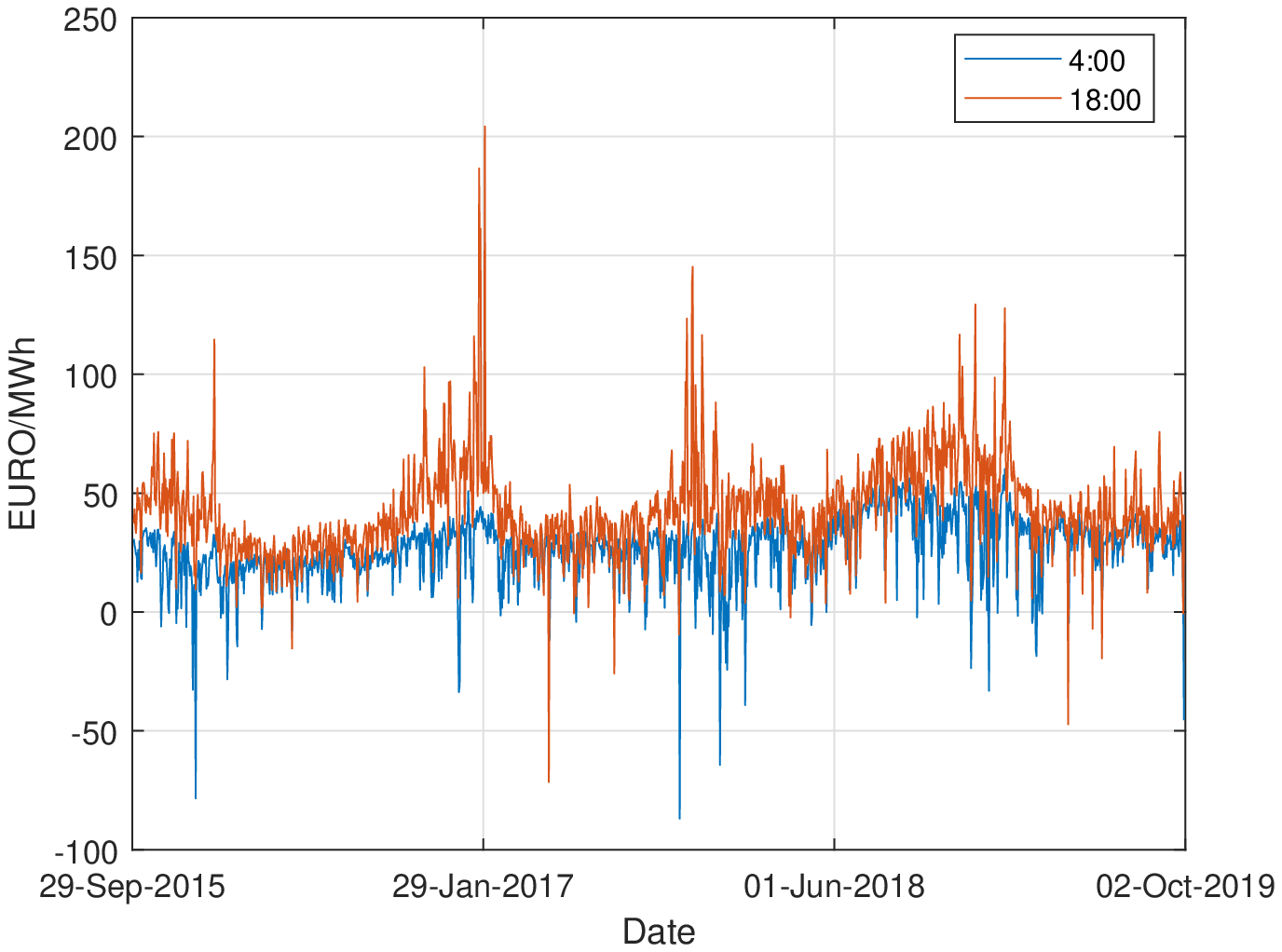}
		\caption{Intraday prices}
	\end{subfigure}
	~
		\begin{subfigure}{0.45\textwidth} 
		\includegraphics[width=\textwidth]{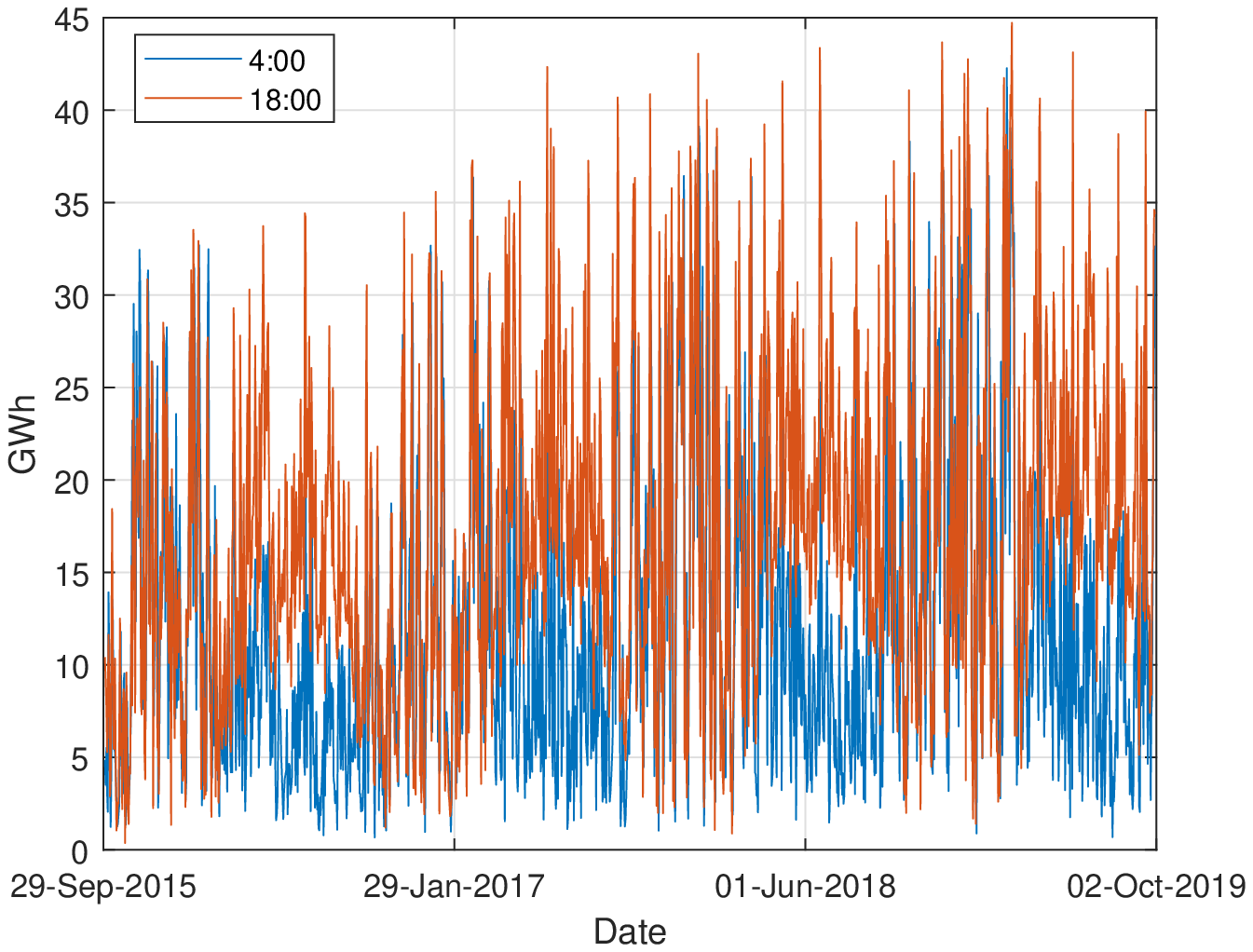}
		\caption{RES}
	\end{subfigure}
	~
	\begin{subfigure}{0.45\textwidth} 
		\includegraphics[width=\textwidth]{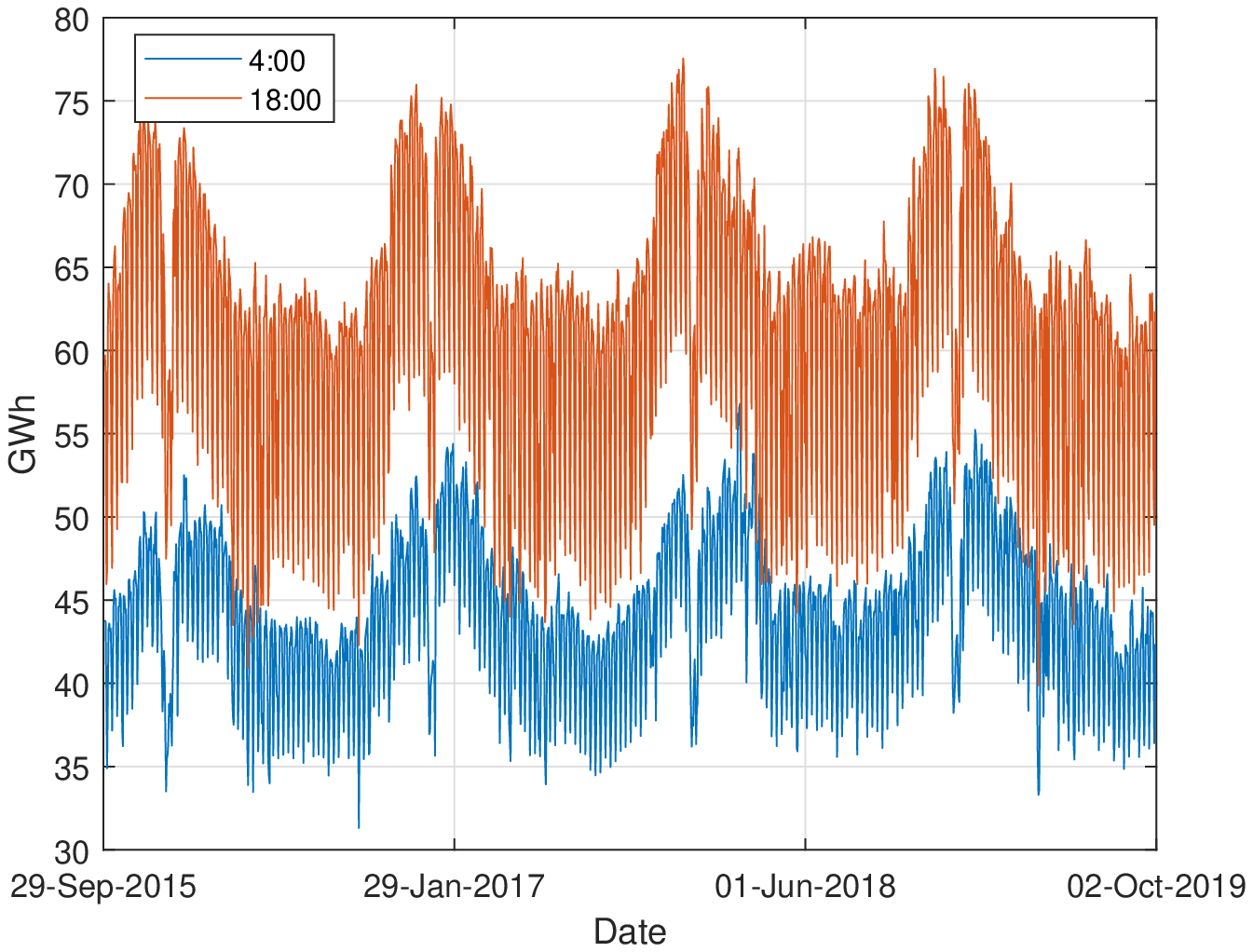}
		\caption{Load}
	\end{subfigure}
\caption{Time paths of electricity prices, RES generation and Load for hours 4:00 and 18:00}
	\label{Fig:data}
\end{figure*}

The properties of times series describing the German electricity market have been extensively studied in the literature. It is well documented that both total load and electricity prices have a strong seasonal pattern. They are on average the highest on working days, during  peak hours. The exemplary time plots of day-ahead and intraday prices together with total load and RES generation are presented in Fig \ref{Fig:data}. Two hours are shown, $h=4,18$, which represent the peak and the off-peak periods of a day. They confirm the  weekly and yearly seasonal behavior of the electricity generation. Additionally, it can be observed that day-ahead and intraday prices co-move together, with the intraday prices being more volatile.  Finally, RES generation reveals different fluctuation pattern, with only minor differences between night and day hours and high variability.

The statistical properties of the data are presented in Table \ref{tab:statistics}. First, the mean and the standard deviation of the variables of interest are computed separately for each hour. Table \ref{tab:statistics} shows their average values across the day. It could be observed that the DA prices are on average slightly lower than ID prices, but also less volatile. Finally, the results of Augmented Dicky-Fuller (ADF) test for the presence of unit roots is presented. The table reports the number of hours, for which the test allows to reject the null and hence confirms the stationarity of the series. It can be observed that majority of analyzed data are stationary. This could be due to a relatively short period of time -- four years -- used for analysis. The result is important for the modeling approach presented below, as it supports an usage of a VAR model for the data in levels.

\begin{table}
	\caption{Statistical properties of the data}
	\label{tab:statistics}
	\centering
	\begin{tabular}{|l|cccc|}
		
		\hline
		& RES & Load & DA & ID \\ 
		\hline
		Mean& 16.40&  61.97&   36.11&   36.24\\
		St. Dev &8.82 &  8.18&   15.61 &  17.06\\
		\hline 
		ADF&	24&    18 &   24&    24	\\
		\hline 
	\end{tabular} 

{\small Notice: ADF indicates the number of hours for which the test rejects the null of an unit root.}		
\end{table}

\section{SVAR model of electricity market}\label{Sec:SVAR}

In this research, VAR model is used to describe the joint behavior of electricity prices and generation, which in turn impact the revenue of the a small RES utility. The literature (see \cite{Weron2014} for a review) indicates that the electricity market has a strong daily seasonality, which impacts not only the level of prices and generation but also its dynamics.  Therefore  each hour is typically modeled separately (see \cite{Ziel:Weron:2018} for a discussion). Here, this implies that the VAR model is built and estimated independently for each hour, $h$. 

In order to capture the major stochastic shocks influencing the variables of interest, a four dimension model is applied:
\begin{equation}
\label{eq:VAR}
Y_{t,h}=A_{0,h}X_{t,h}+\sum_{p=1}^7A_{p,h}Y_{t-p,h} +\varepsilon_{t,h},
\end{equation}
where $Y_{th}=(RES_{t,h}, L_{t,h}, DA_{t,h}, ID_{t,h})'$ is a vector of endogenous variables.
Notice that the vector $Y_{t,h}$ could be extended to include a generation of a particular RES utility, $G_{t,h}$, if such information is accessible. Then, $Y_{t,h}=(G_{t,h}, RES_{t,h}, L_{t,h}, DA_{t,h}, ID_{t,h})'$. Since in this research no such data are available, it is assumed that an exemplary RES producer owns a few wind and solar farms, which are spread across the Germany. For simplicity, it is assumed that $G_{t,h}$ is proportional to the aggregated level of RES generation, with $G_{t,h}=\rho RES_{t,h}$. As the result, $RES_{t,h}$, represents in the model both, the total and the individual generation of RES.

The vector $X_{t,h}$ consists of dummy variables indicating week days, TSO forecasts of RES generation and the total load. Additionally, some nonlinear transformation of lagged prices   (similar to\cite{Uniejewski:etal:2019}, \cite{Ziel:Weron:2018}, \cite{Maciejowska:Uniejewskia:Serafin:2020}) are used: $DA_{min,t-1}$ (a minimum price from the last day), $DA_{max,t-1}$ (a maximum price from the last day) and $DA_{t-1,24}$ (the last known price). The parameters $A_{0,h}$ and $A_{p,h}$ are matrices of dimensions  $(4\times 12)$ and $(4\times 4)$, respectively. They are estimated separately for each hour, $h$.
The residuals, $\varepsilon_{t,h}$, are $(4\times 1)$ random vectors with zero mean and a variance-covariance matrix $\Sigma_h$. Since (\ref{eq:VAR}) is a reduced form of the VAR model, then the residuals  are allowed to be cross-correlated and hence the $\Sigma_h$ is not diagonal.

The reduced form model, although useful for point forecasting of endogenous variables, is not suited for risk analysis -- for example impulse responses or simulations. Therefor, a structural extension of the model (\ref{eq:VAR}) is used.
The SVAR model could take various forms (see \cite{Lutkepohl05}). Here, the B-model is adopted, which focuses on residuals of model (\ref{eq:VAR}). The SVAR  model assumes that the within-sample errors,  $\varepsilon_{t,h}$, are a linear transformation of structural shocks, $u_{t,h}$, which are uncorrelated and have a diagonal variance-covariance matrix, $\Lambda_h$. The relationship between reduced form and structural shocks is described by following equation:
\begin{equation}
\label{eq:Structural_errors}
\varepsilon_{t,h}=B_hu_{t,h}.
\end{equation}
The matrix $B_h$ is called \emph{an instantaneous effect matrix}, because it describes how structural shocks affect endogenous variables in the current time period, $t$. For example, $B_{23,h}$ describes the impact of the third structural shock, $u_{3,th}$ on the second element of $Y_{t,h}$, which is the total load.
Notice that equation (\ref{eq:Structural_errors}) implies that $\Sigma_h=B_h\Lambda_h B_h'$, so there is a direct relationship between $B_h$ and the variance of errors $\varepsilon_{t,h}$. It is typically assumed that either structural shocks have an identity variance-covariance matrix, $\Lambda_h=I$, or the diagonal elements of $B_h$ are equal to one. Here, the first approach is adopted.

Since the structural shocks are assumed to be uncorrelated, which in the Gaussian framework implies independence, their behavior is much easier to model and predict. They could be also simulated separately. Unfortunately, the SVAR model cannot be directly estimated due to lack of identification. As discussed in the literature (see \cite{Lutkepohl05} for a comprehensive discussion on VAR models), the structural model requires estimation of $K^2$ elements of the $B$ matrix, where $K$ is the number of endogenous variables. In the current study, $K=4$, hence the structure is defined by $16$ parameters. At the same time, the variance-covariance matrix $\Sigma$ of the reduced form, due to its symmetry, consists of only $K(K+1)/2=10$ parameters. This implies that there is not enough information to identify the structural parameters. As the result, additional $K(K-1)/2$ assumptions need to be imposed, which will restrict the parameter space. In the presented model, six identification restrictions are need to  ensure model indetifiability. 

In this research, four structural shocks are considered:  weather shock $u_{1,th}$, demand shock $u_{2,th}$, day-ahead speculative shock $u_{3,th}$ and intraday speculative shock $u_{4,th}$. The energy market has its particular features, which helps to recover the structure of SVAR model. First, due to the dispatch priority and support schemes, RES generation does not depend neither on demand nor on price shocks. Second, the literature indicates the limited price elasticity of demand, because market participants require time to adjust their production to the market situation. In particular, the demand response to the unpredicted price innovations is assumed to be insignificant. Finally, the spot pries are set day-ahead, before the intraday trade rises and hence it could be assumed that they do not depend on the intraday speculative shocks.
As the result, the instantaneous effect matrix becomes lower triangular
\begin{equation}
\label{eq:B}
B_h = \left[ \begin{array}{cccc}
* &0&0&0\\
*&*&0&0\\
*&*&*&0\\
*&*&*&*\\
\end{array}\right]
\end{equation}
The zeros in the above matrix correspond to the 'no impact' restrictions. Notice that in the presented solution, the ordering of endogenous variables and structural shocks play an important role. The model implies, for example, that the shock $u_{1,th}$ -- which is the weather shock -- influences all variables, whereas $u_{4,th}$ -- which is the intraday speculative shock -- impacts only intraday prices.

\section{Decision problem of a RES utility}\label{Sec:Decision}

In this article, SVAR model presented in Section \ref{Sec:SVAR} is used for designing trading strategies of a RES utility. Additionally, it is assumed that the generator participates in the day-ahead and the intraday market and is small enough not to impact directly the market prices. It does not receive FIT and hence its revenue is  related to wholesale prices. The utility needs to place an order in the day-head market at noon of the day proceeding the delivery. Due to the stochastic nature of RES, the offered generation differs from the actual production and therefore it needs to trade the difference between the scheduled and the final generation in intraday market. The power plant self-balances its position and therefore no trade is left for the balancing market. 

For illustrative purposes, it is assumed that $G_{th} = \rho RES_{th}$. In order to ensure that RES producer has a minor influence on market prices, the $\rho$ is set to equal $0.5\%$. This implies that in the year 2018, it generated around 85 MWh per hour and accounted for 0.2\% of the total electricity production in Germany.

The utility needs to choose on the day $t-1$,  which part, $g$, of the expected generation, $\hat{G}_{th}$ is offered in the day-ahead market. The remaining part of the production, $G_{th} - g\hat{G}_{th}$, is soled in the intraday market. Notice that in general, the variable $g$ may change across days and hours. Since the utility focuses on the real trade, it is assumed that it does not speculate and hence $g\in[0,1]$. As the result, the revenue from the trade becomes
\begin{equation}\label{eq:pi}
\pi_{t,h}(g)  = g\hat{G}_{t,h}DA_{t,h} + (G_{t,h} - g\hat{G}_{t,h})ID_{t,h}.
\end{equation}
It should be mentioned here that  values of prices and generation are not known at the moment of taking the decision. They depend on stochastic factors, which change throughout the day, such as the weather condition and the human behavior.
As the results, the actual level of revenue, $\pi_{t,h}(g)$ becomes random. It is expected that its distribution is non-normal, as it includes both the level and the product of a few random variables. Moreover, since the generation and electricity price forecast errors are allowed to be correlated, the expected level of revenue will generally by different than $E(\pi_{th}(g))  \neq g\hat{G}_{th}E(DA_{th}) + (E(G_{th}) - g\hat{G}_{th})E(ID_{th}).$

\subsection{Trading strategies}

The utility places its order in DA market on the day preceding the deliver and this transaction cannot be changed, as the new information on  weather conditions and prices arrive. Therefore, its revenue depends directly on the amount of electricity soled there, $g\hat{G}_{t,h}$. As mentioned before, $g$ does not have to be constant and may adjust to the market situation. Therefore in the remaining part of the paper, it will be indexed with a day ($t$) and an hour ($h$): $g_{t,h}$. 

\subsubsection{Simple DA and ID strategies}

First two simple approaches are considered: \emph{day ahead strategy}, which assumes that $g_{t,h}=1$ and \emph{intraday} strategy, for which $g_{t,h}=0$. They are two boundary approaches, which assume a fixed value of the share $g_{t,h}$ that does not depend on market conditions. In the first case, all the predicted production is soled on DA, whereas in the second case the utility decides to wait with the trade till the next day and leave all the generation for the intraday market.

\subsubsection{Data driven strategies}
Next, \emph{data driven strategies} of choosing the level of $g_{t,h}$ are proposed, which utilize the estimated structure of the forecast errors. Using a bootstrap simulation, the optimal proportion of generation offered in DA market is selected to either maximize the expected revenue ($E\pi(g_{t,h})$) or to minimize the risk. Here, the risk is evaluated with two measures: the Sharp Ratio ($SR(g_{t,h})$) and Value-at-Risk ($VaR(g_{t,h})$), which have been shown relevant and useful by the financial literature.

In this research, a bootstrap procedure is proposed to generate the distribution of $Y_{t,h}$. This allows to approximate the distribution of the revenue for different levels of $g$ and to optimize its value. The algorithm consists of the following steps
\begin{enumerate}
	\item For a selected hour $h$ and the calibration window $\{t_0,t_0+1,...,t-1\}$ estimate the parameters of SVAR model: $ \hat{A}_{p,h}$ for $p=0,1,...,7$ and the instantaneous effect matrix $\hat{B}_h$
	\item Calculate the point forecasts of $\hat{Y}_{t,h}$ according to (\ref{eq:VAR}) as 
	\begin{equation}\label{eq:Y_forecast}
	\hat{Y}_{t,h} = \hat{A}_{0,h}X_{t,h} + \sum_{p=1}^7\hat{A}_{p,h}Y_{t-p,h}
	\end{equation}
	and $\hat{G}_{t,h}$ as $\hat{G}_{t,h} = \rho \hat{RES}=\rho\hat{Y}_{1,th}$.
	\item Compute residuals $\hat{\varepsilon}$ of a reduced form model (\ref{eq:VAR}) and corresponding structural shocks $\hat{u} = \hat{B}_h^{-1}\hat{\varepsilon}$
	\item Approximate the distribution of the next day $Y_{t,h}$ and profits $\pi_{t,h}(g)$ using a bootstrap sampling of structural shocks. For each iteration $b=1,...,B$
	\begin{itemize}
		\item Pick independently a realization of each shock and obtain a $(4\times 1)$ vector $\tilde{u}^{(b)}$
		\item Transform structural shocks into forecast errors:  $\tilde{\varepsilon}^{(b)}=\hat{B}_h\tilde{u}^{(b)}$
		\item Calculate $\tilde{Y}_{t,h}^{(b)}=\hat{Y}_{t,h}+\tilde{\varepsilon}^{(b)}$, where $\tilde{Y}_{t,h}^{(b)} = ( \tilde{RES}^{(b)}_{t,h}, \tilde{L}^{(b)}_{t,h}, \tilde{DA}^{(b)}_{t,h}, \tilde{ID}^{(b)}_{t,h})'$, and $\tilde{G}_{t,h}^{(b)}=\rho \tilde{RES}^{(b)}_{t,h}$ 
		\item For the chosen level of $g$ compute the corresponding profit $\tilde{\pi}_{t,h}^{(b)}(g)$ according to (\ref{eq:pi})
	\end{itemize}
		
	\item Estimate the expected value and the variance of incomes, $E\pi_{t,h} (g)$ and $\sigma_{t,h}^2(g)$ as the mean and the mean squared deviation of $\tilde{\pi}_{t,h}^{(b)}(g)$ across $b$. Using this information, compute the Sharp ratio as $SR_{t,h}(g) = E\pi_{t,h}(g)/\sigma_{t,h}(g)$ and approximate $VaR_{t,h}^{\tau}(p)$ by a $\tau$ quantile of $\tilde{\pi}^{(b)}_{t,h}(g)$
	
\end{enumerate}
The selection of the level of $g_{t,h}$ depends on the optimality condition considered.
The first approach picks $g_{t,h}^*$, which maximizes the expected revenue $E\pi_{t,h}(g)$. This method, although profitable, may result in an increase of the transaction risk. Therefore, alternative trading strategies are also examined, which aim at minimizing the risk by choosing $g_{t,h}^*=argmax_g{SR_{t,h}(g)} $ or $g_{t,h}^*=argmax_g{VaR_{t,h}^{\tau}(g)} $. The brief description of the approaches together with their notation are presented in Table \ref{tab:strategy:notation}.

\begin{table}
	\caption{Trading strategies}
	\label{tab:strategy:notation}
	\centering
	\begin{tabular}{|l|c|c|c|}	
		\hline
		
		Name& Description & Optimality criteria & $g$ \\ 
		\hline
		$S_{DA}$& Sell the expected generation on DA market& - & 1\\
		$S_{ID}$& Sell all the generation on Intraday market &  - &0\\
		$S_{E\pi}$& Maximize expected profit &  $\max E\pi(g)$  & $g^*_{t,h}$\\
		$S_{SR}$& Maximize Sharp ratio  &  $\max SR(g)$  & $g^*_{t,h}$\\
		$S_{VaR}$& Maximize Value of Risk for the $5\%$ quantile &  $\max VaR_{0.05}(g)$ &$g^*_{t,h}$\\		
		\hline 		
	\end{tabular} 			
\end{table}

\subsection{Evaluation of trading strategies}

The trading strategies presented above can be compared according to various dimensions. Here, three features are used to evaluate their performance: the level of revenue, its predictability and variability.

\subsubsection{The level of income}
In this article, the average hourly revenue is used to analyze the performance of presented approaches. It is computed as follows:
\begin{equation}
\label{eq:revenue}
\bar{\pi} = \frac{1}{T_{eval}\times 24}\sum_{t,h} \pi_{t,h}(g_{t,h}),
\end{equation} 
where $T_{eval}$ is the number of days used for evaluation.
In order to verify, if a chosen strategy, $i$, yields a higher average income than the starategy  $j$, a new variable is defined
$$ d_{t,h} = \pi_{t,h}(g^{(i)}_{t,h}) - \pi_{t,h}(g^{(j)}_{t,h}).$$
If both strategies are characterized by the same expected value then $Ed_{t,h}=0$. Hence, the natural hypothesis in this setup are $H_0:Ed_{t,h}=0$ and $H_1:Ed_{t,h}>0$. Under the null, both strategies provide the same average revenue, whereas under the alternative, the strategy $i$ is more profitable than $j$. In order to verify the hypothesis, a \cite{dieb:mar:1995} type testing procedure is applied to the mean daily level of $d_{t,h}$: $d_t=1/24\sum_{h=1}^{24}d_{t,h}$. The variable  $d_t$ can be viewed as a counterpart of the loss differential, for example a difference of squared forecast errors. The statistic takes the form
\begin{equation}\label{eq:DM}
DM = \frac{\bar{d}}{\sqrt{2\pi \hat{f}_d(0)/T_{eval}}},
\end{equation}
where $\bar{d}$ is the average value of $d_t$ and $\hat{f}_d(0)$ is an estimator of spectral density of $d_t$ at frequency 0. As $2\pi \hat{f}_d(0)/T_{eval}$ is an estimator of the variance of $\bar{d}$, $DM$ converges asymptotically to a standard normal distribution, $N(0,1)$. 

\subsubsection{Risk}

The risk associated with a strategy is evaluated according to two features: possibility to predict accurately the next day revenue and the variability of the  income. In order to asses the revenue forecast quality, we use the outcomes of a SVAR model. The revenue predictions are calculated as the average of $\tilde{\pi}^{(b)}(g)$ across bootstrap iterations:
\begin{equation}
\hat{\pi}_{t,h} = \frac{1}{B}\sum_{b=1}^{B}\tilde{\pi}^{(b)}_{t,h}(g_{t,h})
\end{equation}
First, it is checked, how much the predicted revenue differs from its actual values, $\pi_{t,h}(g_{t,h})$. In this research, similar to other EFP papers, the forecast quality is evaluated with Root Mean Squared Error (RMSE) and Mean Absolute Error (MAE)
\begin{equation}
\label{eq:RMSE}
RMSE = \sqrt{\frac{1}{T_{eval}\times 24}\sum_{t,h} (\pi_{t,h}-\hat{\pi}_{t,h})^2}.
\end{equation}

\begin{equation}
\label{eq:RMSE}
MAE = \frac{1}{T_{eval}\times 24}\sum_{t,h}|\pi_{t,h}-\hat{\pi}_{t,h}|.
\end{equation}
In order to statistically verify if forecasts stemming from two different strategies, $i$ nad $j$, are equally accurate, the DM test is applied to the following loss differentials:
$$ d_{t} = \frac{1}{24}\sum_{h=1}^{24}{(u_{t,h}^2(i)-u_{t,h}^2(j))}$$
for RMSE and
$$ d_{t} = \frac{1}{24}\sum_{h=1}^{24}{|u_{t,h}(i)-u_{t,h}(j)|}$$
for MAE measure. The $u_{t,h}(i)$ and $u_{t,h}(j)$ are forecast errors of analyzed strategies: $u_{t,h}(i) =\pi_{t,h}(g^{(i)}_{t,h})-\hat{\pi}_{t,h}(g^{(i)}_{t,h}) $ and $u_{t,h}(j) =\pi_{t,h}(g^{(j)}_{t,h})-\hat{\pi}_{t,h}(g^{(j)}_{t,h}) $. The test statistics have a form analogous to (\ref{eq:DM}).

Finally, the risk associated with the variability of income is measured by the Value-at-Risk of revenues for a given hour. In order to aggregate the results, the average VaR is used to compare the outcomes of different strategies.

\section{Results}\label{sec:Results}

\subsection{Experiment design}
In order to asses different trading strategies, a forecasting experiment is run, in which a moving window methodology is applied. In the experiment, the calibration window includes 731 observations and the results are assessed with the last two years from 01-Oct-2017 to 31-Sep-2019 ($T_{eval}=730$ observations). Moreover, in order to bring the experiment as close as possible to the empirical problem, different information sets are used for hours before and after the time of taking the decision. In this work it is assumed that the generator places on order on DA market at 12:00. Therefore, when performing predictions it knows the actual generation and intraday  prices only for hours from midnight till 10:00. For the remaining hours, the actual values ($L_{t-1,h}$, $RES_{t-1,h}$) are not known and are replaced by their TSO predicted levels (see \cite{Maciejowska:Nitka:Weron:2021} for more details). In case of $ID_{t-1,h}$, which is also unavailable, it is assumed that is has no impact on endogenous variable  and hence the last column of $A_{1,h}$ matrix is set to equal zeor. 
Finally, following the energy forecasting literature (\cite{Ziel:2016}, \cite{Uniejewski:etal:2019}, \cite{Maciejowska:Uniejewskia:Serafin:2020}), three lags are selected in the VAR model: $p=1,2,7$.

\subsection{Comparison of trading strategies}

Let us first look at the aggregated results for each trading strategy. The outcomes are reported in Table \ref{tab:strategy:profits}, in which the second column presents the average hourly revenue, $\bar{\pi}$. The remaining three columns focus on risk that is measured by $RMSE$ and $MAE$ of revenue forecasts and $VaR_{1\%}$ and $VaR_{5\%}$ of income. 
It should be noticed that the benchmark results of $S_{DA}$ strategy are expressed in nominal values. The remaining outcomes are presented relative to the benchmark, as the percentage difference ($\%\Delta$). Hence for all the  strategies, apart from $S_{DA}$, values lower than zero indicate that the given strategy reduces the indices and larger then zero prove the opposite.

The analysis of simple strategies, which assume a constant level of $g_{t,h}$, confirm that the transactions made in ID market are slightly more profitable then in DA market. However, the rise of the average revenue in $S_{ID}$ is achieved at the cost of a higher risk. The revenue forecasts suffer from the loss of accuracy (RMSE increases by 9\% and MAE by 7.5\% as compared to $S_{DA}$). Moreover, the income itself is substantially more volatile -- the  $VaR_{1\%}$ drops by almost then 40\%.

At the same time, the data driven approaches provide results characterized by a higher income and lower risk than the benchmark. As expected, the $S_{E\pi}$ strategy allows to earn the largest revenue among all the strategies. It reduces RMSE but its impact on risk is mixed: it lowers $VaR_{1\%}$ and increases $VaR_{5\%}$. The $S_{VaR}$ strategy, which aims at maximizing the $VaR_{5\%}$ of revenue, leads to a rise of income by less than 1\%, which is the weakest result among the data driven approaches. It provides more accurate predictions than the simple $DA$ strategy and reduces RMSE and MAE by more than 6\%. Similar to $S_{E\pi}$ case, its impact on risk is ambiguous.

Finally, the strategy based on the Sharp Ratio provides the most promising results. It allows to increase the revenue and at the same time reduces the risk measure by both RMSE and VaR. The decrease of uncertainty is substantial, as it reduces RMSE and MAE by  8\% and 6.7\%, respectively. Moreover, it increases 1\% and 5\% VaR by 19.5\% and 10.3\%. 

\begin{figure}[h]	
	\centering
	\vspace{1ex}
	\includegraphics[width=0.45\textwidth]{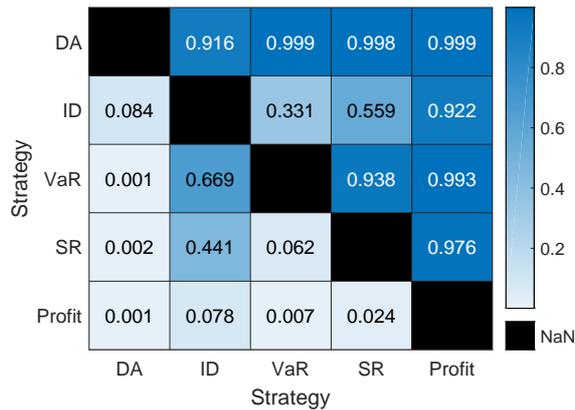}	
	\caption{Results of equal profitability test: $p$-values}	
	\label{Fig:heatmap:profit}
\end{figure}

\begin{figure*}[h]
	\centering	
	\begin{subfigure}{0.45\textwidth} 
		\includegraphics[width=\textwidth]{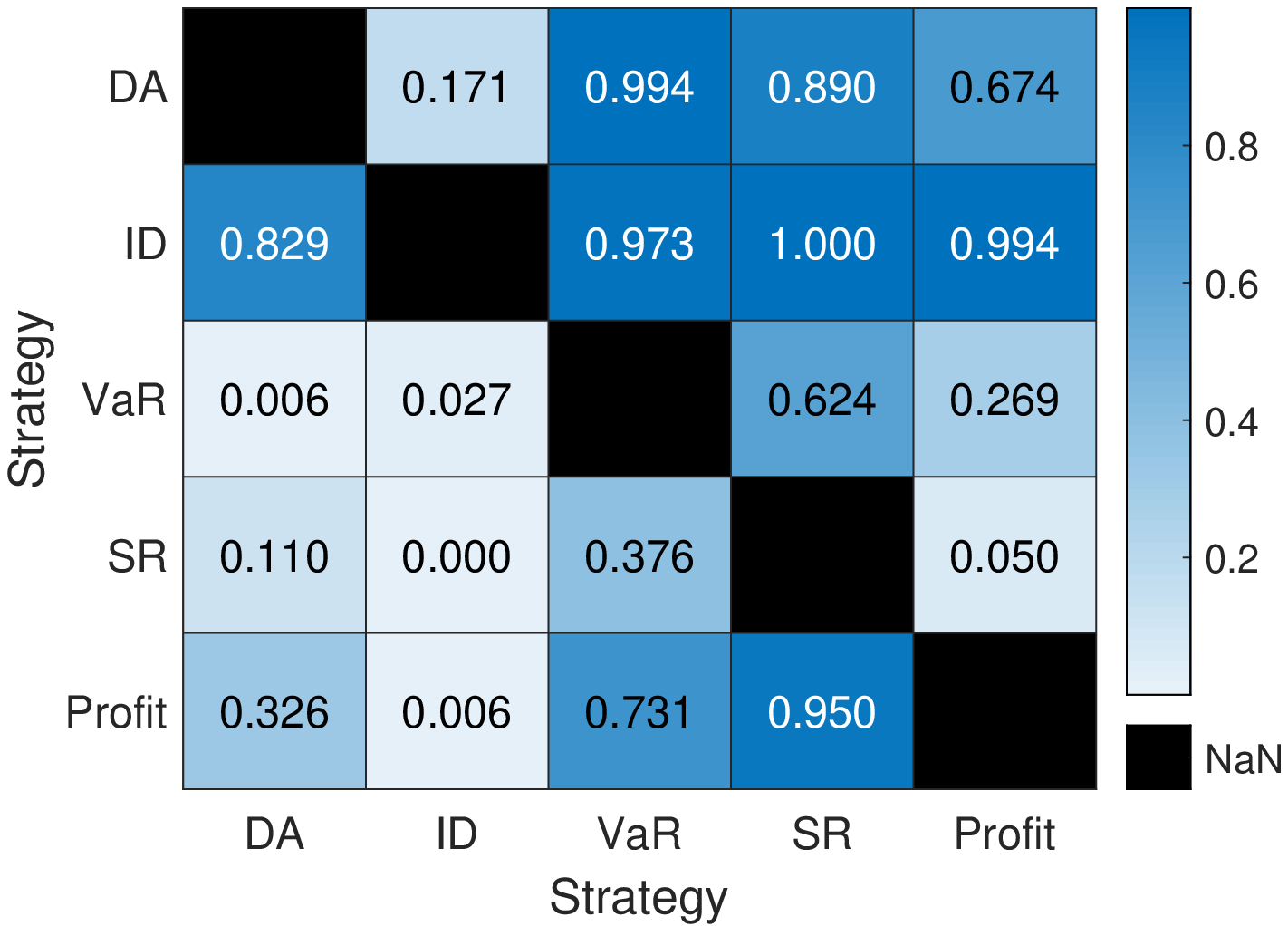}
		\caption{RMSE}
	\end{subfigure}
	~
	\begin{subfigure}{0.45\textwidth} 
		\includegraphics[width=\textwidth]{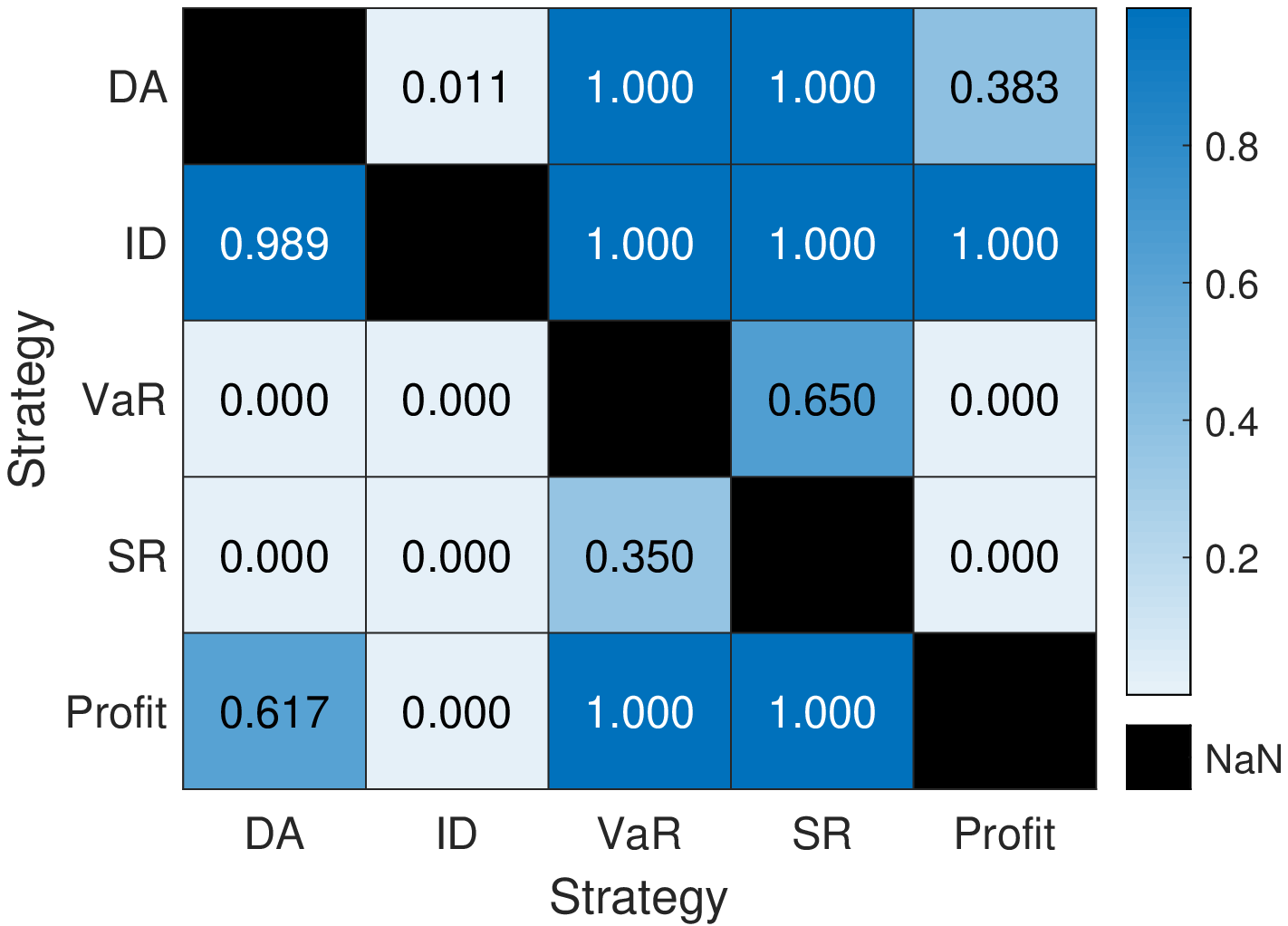}
		\caption{MAE}
	\end{subfigure}
\caption{Results of equal forecast accuracy test for two loss functions(RMSE, MAE): $p$-values}
	\label{Fig:heatmap:forecast}
\end{figure*}

Fig. \ref{Fig:heatmap:profit} presents the asymptotic $p$-values of test statistics used to compare the profitability and Fig. \ref{Fig:heatmap:forecast} depicts comparison of RMSE (left panel) and MAE (right panel) forecast accuracy measures. It should be recalled that if the $p$-value of $DM_{i,j}$ statistic is smaller than 10\%, then it implied that the strategy $i$ provides a higher revenue or more accurate predictions than the strategy $j$. The results confirm previous outcomes that all strategies provide higher income than the benchmark, $S_{DA}$. Moreover, $S_{E\pi}$ (denoted on plots as Profit), brings the largest revenue among the presented approaches. When the RMSE and MAE of forecast errors are considered, the outcomes show that $S_{DA}$ allows to predict the next day revenue more accurately than $S_{ID}$, which is the worst strategy according to this measure. Finally, the data driven methods aiming at minimizing the risk leads to a significant reduction of both forecast accuracu measures.

To sum up, the aggregate results indicate that using bootstrap method based on SVAR model for forecasting generation and constructing trading strategies could bring additional profit and at the same time reduce the risk. Hence, it  it is preferable from the perspective of a small RES utility. Among presented approaches, $S_{SR}$ is the most attractive one. It does not only increase substantially VaR but brings on average 41.16 EURO more revenue per hour than the benchmark, which is equivalent to an increase of profit within two years by 721 117.59 EURO.

\begin{table}
	\caption{The average hourly revenue and risk measures}
	\label{tab:strategy:profits}
	\centering
	\begin{tabular}{|c|c|cc|cc|}	
		\hline
		Strategy& $\bar{\pi}$ & $RMSE$& $MAE$&$VaR_{1\%}$ & $VaR_{5\%}$ \\ 
		\hline
		$S_{DA}$& 3215.6 &1196.0&635.5 &-1803.1& 758.7\\
		\hline
		& \multicolumn{5}{|c|}{$\%\Delta$} \\
		\hline
		$S_{ID}$& 1.20 & 8.95 & 7.46 & -39.96 & -1.04\\
		$S_{E\pi}$& 1.84  &-3.24&0.66 &-4.13 & 7.07\\
		$S_{SR}$& 1.28 & -8.00&-6.67 & 19.51 & 10.29 \\
		$S_{VaR}$& 0.91 & -6.50 & -6.29 & -4.44 &  6.09\\		
		\hline 		
	\end{tabular} 			
\end{table} 

\subsection{Distribution of $g$}

\begin{table}
	\caption{The averaged proportion of generation offered on DA market: \% points}
	\label{tab:strategy:g}
	\centering
	\begin{tabular}{|l|c|ccc|}
		\hline
		Strategy&  $\bar{g}$ & $g=0$ & $0<g<1$  & $g=1$\\ 
		\hline
		$S_{DA}$& 100.0& 0.000&0.000& 100.0\\
		$S_{ID}$& 0.000& 100.0&0.000& 0.000\\
		\hline
		$S_{E\pi}$&  49.05& 50.95&0.000 &49.05\\
		$S_{SR}$& 51.36& 12.47& 84.46&3.08\\
		$S_{VaR}$& 52.72& 9.68& 85.84&4.48\\			
		\hline 		
	\end{tabular} 			
\end{table} 

Let us now look more into details of the presented outcomes. First, the data driven approaches allow to choose a proportion of predicted generation, which is offered in DA market. Table  \ref{tab:strategy:g} shows the average level of optimal $g_{t,h}$ for different strategies ($\bar{g}$) together with the proportion of its values in three groups: $g=0$, $0<g<1$ and $g=1$. All the quantities are expressed in \% points.

It is clear that simple strategies provide two boundary outcomes with $\bar{g}=1$ or $\bar{g}=0$. This is due to the fact that for these two approaches all the predicted generation is offered either in DA or ID market. The more complex results are observed for the data driven strategies, for which the average level of $g$ oscillates around 50\%. First, it can be noticed that in case of $S_{E\pi}$ strategy  $g_{t,h}$ takes only the extreme values:  $0$ or $1$. It selects $ID$ market in 50.95\%, which confirms the previous findings that indicate larger profitability of this market. 

\begin{figure}
	\centering	
	\vspace{1ex}
	\includegraphics[width=0.6\columnwidth]{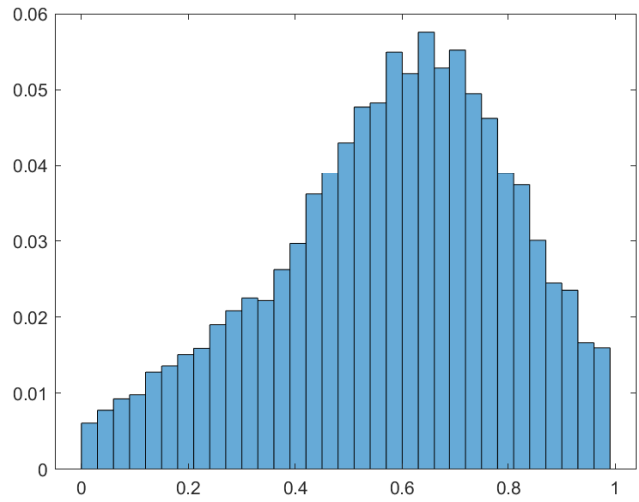}
	\caption{Frequences of $g$ values such that $0<g_{t,h}<1$ for strategy $S_{SR}$ }		
	\label{Fig:hist_g}
\end{figure}

When the strategies aiming at minimizing the risk, $S_{SR}$ and $S_{VaR}$, are considered, it can be observed that the algorithm selects $g_{t,h}=0$ and $g_{t,h}=1$ in around 11\% and 4\% cases, respectively. Hence, for around 85\% of hours, the proportion of generation soled in DA market falls into the interval $(0,1)$. The distribution of $0<g_{t,h}<1$ for $S_{SR}$ is depicted in Fig. \ref{Fig:hist_g}. The plot shows that in majority of cases,  $g_{t,h}$ takes the value between $0.5$ and $0.8$. Moreover, its distribution is skewed to the right indicating that the DA market is more attractive than the ID one. 

\begin{figure}	
	\centering	
	\includegraphics[width=0.6\columnwidth]{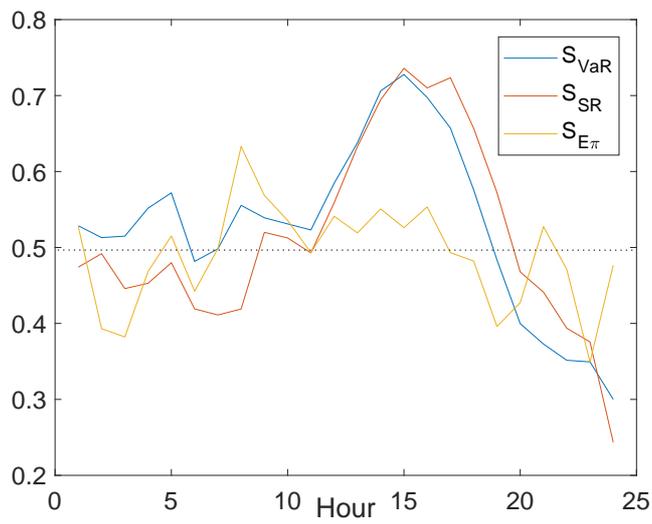}	
	\caption{Average value of $g_{t,h}$ across hours}
	\label{Fig:g:hours}	
\end{figure}

The average levels of $g_{t,h}$ across 24 hours is presented in Fig. \ref{Fig:g:hours}. The most risky, $S_{E\pi}$ strategy is characterized by lower values of $g$ and hence tends to sell a larger share of the generation in ID market than $S_{VaR}$ and $S_{SR}$. When these two are considered, it can be observed that the highest average value of $g$ is obtained for hours 15-17, when on average more than 70\% of predicted generation is offered in DA market. On the contrary, during the evening hours, from 19-24, these strategies suggest selling majority of the production in the ID market. For night and early morning hours the results are mixed, as the average value of $g$ is close to 50\%.

\section{Conclusions}\label{Sec:Conclusions}

The changes in the electricity markets expose RES generators to various risks, among which the price and the volume risk play a very important role. RES generators, which revenue depends on the market prices and the offered quantity, can now actively build a portfolio from different types of contracts. In this research, it is assumed that it trades produced electricity either in day-ahead or intraday market. It can be noticed that due to an intermittent generation and stochastic electricity prices, the entrepreneur acts under strong uncertainty.

In this research, it is assumed that the trading portfolio constructed by the utility depends on two major factors: predicted level of production and the chosen share of generation, $g_{t,h}$, offered in the DA market. Since the utility does not speculate, it is assume that $g_{t,h}$ belongs to an interval $(0,1)$. As a result, two types of trading strategies are considered: simple strategies -- which assume a fixed value of $g_{t,h}$ -- and data driven strategies. In the later one, the SVAR model is used to predict future level of generation and to select an optimal level of $g_{t,h}$ in order to either maximize the revenue or minimize the transaction risk. 

The performance of presented trading strategies is next compared using the data from German electricity market. The results of the research indicate that the transactions in ID market are on average more profitable than in DA market. They are however burden with a significantly larger risk. Second, the data driven strategies provide revenues larger than the benchmark, $S_{DA}$, strategy and at the same time allow to reduce risk measured by the predictability (RMSE, MAE) of a next day revenue and its variability ($VaR$). Among the proposed approaches, the strategy maximizing the Sharp Ratio is the most promising one, as it provides the most robust outcomes.

Finally, the selected shares of predicted generation offered in DA market, $g_{t,h}$, are analyzed. The results show that the strategy aiming at maximizing the profit, chooses only the two boundary values of $g_{t,h}$ and hence offers all the forecasted production in either DA or ID market. On the contrary, approaches minimizing the risk, are more prone to build the portfolios from both markets at the same time.

This research provides a promising approach for constructing trading portfolio of a small RES producer. This comprehensive method allows to explore information on different aspects of the market: the structure of generation and electricity prices. The results indicate that using such diversified information set can lead to a significant increase of profits and a reduction of a transaction risk. Finally, the proposed approach can be further extended to allow for other types of contracts or trading strategies.

\section*{Acknowledgments}
This work was partially supported through SONATA BIS grant no. 2019/34/E/HS4/00060 and SONATA grant no. 2015/17/D/HS4/00716.

\bibliographystyle{elsarticle-harv}
\bibliography{bibliography}

\end{document}